\documentclass[12pt]{article}
\usepackage[pctex32]{graphics}
\begin{document}
~
~
\vspace{1cm}
\begin{center} {\Large \bf  Anti-de Sitter D-branes in Curved Backgrounds}
                                                  
\vspace{1cm}

                      Wung-Hong Huang\\
                       Department of Physics\\
                       National Cheng Kung University\\
                       Tainan, Taiwan\\

\end{center}
\vspace{1cm}
\begin{center} {\large \bf  Abstract} \end{center}
We investigate the properties of the AdS D1-branes which are the bound states of a curved D1-brane with $n$ fundamental strings (F1) in the $AdS_3$ spacetime, and the AdS D2-branes which are the axially symmetric bound states of a curved  D2-brane with $m$ D0-branes and $n$ fundamental strings in the $AdS_3 \times S^3$ spacetime.  We see that, while the AdS D1-branes asymptotically approach to the event horizon of the $AdS_3$ spacetime the AdS D2-branes will end on it.  As the near horizon geometry of the F1/NS5  becomes the spacetime of $AdS_3 \times S^3 \times T^4$ with NS-NS three form turned on, we furthermore investigate the corresponding AdS D-branes in the NS5-branes and macroscopic F-strings backgrounds, as an attempt to understand the origin of the AdS D-branes.   From the found DBI solutions we see that in the F-strings background, both of  the AdS D1-branes and AdS D2-branes will asymptotically approach to the position with a finite distance away from the F-strings.   However, the AdS D2-branes therein could also end on the F-strings once it carries sufficient D0-branes charges.  We  also see that there does not exist any stable AdS D-branes in the NS5-branes backgrounds.  We present physical arguments to explain these results, which could help us in understanding the intriguing mechanics of the formation of the AdS D-branes.

\vspace{2cm}
\begin{flushleft}
E-mail:  whhwung@mail.ncku.edu.tw\\
\end{flushleft}
\newpage
\section {Introduction}
The Anti-de Sitter spacetimes ($AdS_n$) are known to play important roles in the evolution of the string theory.  The famous AdS/CFT correspondence relates the quantum gravity in the $AdS_n$ to a dual conformal field theory living on the boundary [1].   The spacetime of  $AdS_3 \times S^3 \times T^4 $ with NS-NS gauge fields turned on arises from the embedding of a stack of fundamental strings within a stack of NS-5-branes, as one takes the near-horizon limit.   The case of $AdS_3$ together with NS-NS three form provides an exact string background, which is described  in terms of SL(2,{\bf R}) Wess-Zumino-Witten (WZW) model [2].   It provides a unique setting in which to analyze the AdS/CFT correspondence beyond the supergravity approximation.  The study of D-branes  in the AdS backgrounds is also a problem of significant interest and have been extensively studied [3-10].  

   The WZW D-branes for compact Lie groups have been by now rather well understood both from the conformal field theory (CFT)  and from the geometric, target-space viewpoint [3].  For the non-compact group  $SL(2,\bf R)$, the  analysis by Bachas and Petropoulos [4] have found a stable $AdS_2$ branes which are the curved bound states of a D1-brane with $n$ fundamental strings (F1).  These $AdS_2$ branes reveal some interesting features, such as the quantization conditions, that were not presented in the  case of compact groups.  The problems of constructing the boundary states and supergravity solutions for the D-branes in the $AdS_3$ background had also been studied by several authors [5-10].

   In this paper we will use the Dirac-Born-Infeld action to find AdS D-branes solutions in the $AdS_3 \times S^3$ and macroscopic fundamental strings  backgrounds.   The first class  is called as the AdS D1-branes (see the figure 1) which are the bound states of a curved D1-brane with $n$ fundamental strings. (It is called as $AdS_2$ brane in the convention [4-10].)  We will see that there are two types of AdS D1-branes.  The first will asymptotically approach to the position with a finite distance away from the event horizon (i.e. the value of  ``d" in figure 1 is not zero) while the second will approach to the event horizon asymptotically.
\\
\vspace{1cm}
\hfil\scalebox{1}{\includegraphics{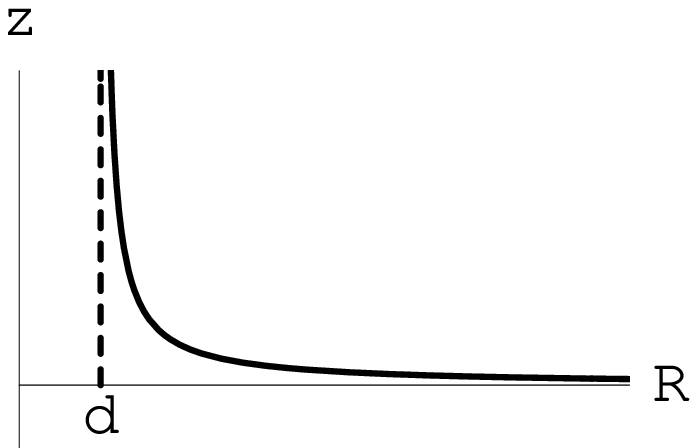}}\hfil\\
{\it ~~~Figure 1. The profile of the AdS D1-branes which are the bound states of a D1-brane with $n$ fundamental strings in macroscopic string background.}
\vspace{1cm}
\\
The second class  is called as the AdS D2-branes which are the axially symmetric bound states of a curved D2-brane with $m$ D0-branes and $n$ fundamental strings.  We will see that there are three types of AdS D2-branes.    The first will asymptotically approach to a position with a finite distance away from the event horizon while the second will approach to the event horizon asymptotically (see the figure 2 plotted in section 2-2).   The third type of AdS D2-branes will end on the event horizon (see the figure 3 plotted in section 2-2). 

  Our motivations are to understand the original mechanics of forming these AdS D-branes.  For example, how the curvature of spacetime will render a straight D1-brane to  a curved  AdS D1-brane as that shown in the figure 1?  What effect could  make a AdS D2-brane to posses the curved geometry likes that shown in figure 2 instead of a tubular configuration [11-12]?  To arrive our goal it is worth to notice that the anti-de Sitter spacetime arises naturally in some limits of the stringy spacetimes.   For example, the near-horizon geometry of the F1/NS5 system becomes the spacetime of $AdS_3 \times S^3 \times T^4$ with NS-NS three form.   Therefore the  investigations of  the corresponding AdS D-branes in the NS5-branes and macroscopic fundamental strings backgrounds could help us in finding the physical reasons of forming these AdS D-branes.   

    In section 2, after reviewing the AdS D1-branes found in [4] we begin to analyze AdS D2-branes in the $AdS_3 \times S^3$ spacetime.  We see that, while the AdS D1-branes asymptotically approach to the event horizon of $AdS_3$ the AdS D2-branes will end on it. In section 3, we  show that there does not exist any corresponding stable AdS D-branes in the NS5-branes background.  In section 4 we analyze the AdS D1-branes and AdS D2-branes in the F-strings background.  We see that, both of  the AdS D1-branes and AdS D2-branes will keep a finite distance away from the F-strings.   However, the AdS D2-branes could end on the F-strings once it carries sufficient large D0-brane charges.    We present physical arguments in each section to explain these results, which could help us to understand the intriguing mechanics of the formation of the AdS D-branes.    We make a conclusion in the last section.

  Note that in recent many authors have investigated the dynamics of D-branes in the nontrivial background, including the NS5-branes [13,14], BPS Dp-branes [15] and macroscopic fundamental strings [16] backgrounds.  However, their results are used to show that the radial mode of the BPS D-brane in these backgrounds resembles the tachyon rolling dynamics of unstable D-brane [17], and have not discussed the AdS D-branes therein.   Note also that the possible tubular solution in these background had been discussed by us in [12].   The present works are the complements of [12-16] and could further our understanding of the D-branes in the curved stringy backgrounds.

\section {AdS D-branes in $AdS_3 \times S^3$ Background}
\subsection {AdS D1-branes in $AdS_3$ Background}
The background metric and NS-NS potential in the $AdS_3 $ background  is 
$$ds^2= - r^2 \, dt^2 + r^2\,dz^2 + r^{-2}\,dr^2,\eqno{(2.1a)}$$ 
$$B^{NS}= ~r^2 \, dt \,\wedge \,dz , ~~ \hspace{2.5cm}\eqno{(2.1b)}$$ 
which are described in the Poincar{\'e} coordinates [4].  The Dirac-Born-Infeld action for a D-string in the $AdS_3 $ background  is
$${\cal S} =  - \int dt\, dz \, \sqrt{- \det (g +  F + B^{NS})}\,, ~~~~ F = E\, dt \,\wedge \,dz, \eqno{(2.2)} $$
where $g$ is the induced worldvolume 2-metric, $F$ is the DBI 2-form
field strength which has only a nonzero component $F_{0z}$ [4].   Substituting the metric of the background and the relevant fields into (2.2) we can find the Lagrangian of a static D-string  
$${\cal L} = - \sqrt{R^4 + R'^2 - (R^2 + E)^2},\eqno{(2.3)} $$
in which  $``R"$ describing the position of the D-string and is a function of $z$.    To proceed, we first define the momentum conjugate to the DBI electric field $E$ [4]
$$\Pi \equiv {\partial{\cal L}\over \partial E} = {R^2 + E \over \sqrt{R^4 + R'^2 -(R^2 + E)^2}},   \eqno{(2.4)}$$ 
which implies that 
$$ E = {\Pi\over \sqrt{1+ \Pi^2}~}\,\sqrt{R^4 + R'^2} - R^2 .\eqno{(2.5)}$$ 
The corresponding energy density of the AdS D1-brane is thus
$$ {\cal H} \equiv \Pi E - {\cal L} =  \sqrt{1+\Pi^2}\, \sqrt{R^4+ {R^\prime}^2 } - \Pi \, R^2 .  \eqno{(2.6)}$$
To  solve  the associated $``R(z)"$ equation Bachas and Petropoulos [4] have used the continuity equation of the energy-momentum tensor.  Because the Lagrangian (2.3) does not explicitly depend on the variable $z$ there is two-dimensional Poincar{\'e} invariance, thus the corresponding tensor is conserved. The energy-momentum tensors are defined by 
$$\Theta ^\alpha_{\; \beta}=  {\partial {\cal L} \over \partial  \partial_{\alpha}u} \,  \partial_{\beta} R + {\partial {\cal L}\over \partial F_{\alpha\gamma}}\, F_{\beta\gamma}  - \delta^\alpha_{\; \beta}\, {\cal L}\, ,    \eqno{(2.7)}$$ 
where the Greek indices run over $(t,z)$.  So that in our system, by the continuity equation, $\Theta^z_{\;z}$ is  a world-sheet constant.  A straightforward calculation gives:
$$\Theta^z_{\; z} = { \sqrt{1+\Pi^2}\, \, R^{4} \over \sqrt{R^4+ {R^\prime}^2 }} - \Pi \, R^2 .      \eqno{(2.8)}$$ 
Suppose  that $z$-momentum does not flow out of the string at infinity - this amounts to free boundary conditions in the direction of the event horizon, we  must then demand that  $\Theta ^z_{\; z}=0$, which is possible only for $\Pi \ge 0$, and it has the general solutions 
$$ R(z) =  {|\Pi|\over z-z_0}.      \eqno{(2.9)}$$ 
This is the AdS$_2$ branes found in [4].  We will called it as AdS D1-brane as it is the bound state of (1,$\Pi$)-string in AdS spacetime.  Some comments about the properties of AdS D1-brane are as follows [4]: 

   $\bullet$ The constant $ \Pi$ which determines the radius of AdS$_2$, is  proportional to the number of fundamental strings on the AdS D-brane.  The radius of AdS$_2$  is thus quantized in units of the string coupling constant. 

   $\bullet$  Substituting the solution (2.9) into (2.5) we see that $ E = 0$, which tell us that the  gauge-invariant meaning can, however, only be attached to ${\cal F}_{tz} = B^{NS} + F$.   

   $\bullet$  The fact that $\Pi$ shall be positive to have a AdS D1-brane implies a definite orientation for the fundamental strings on it.

    $\bullet$ The corresponding energy density of the AdS D1-brane is 
$$ {\cal H}_{AdS D1-brane}  = {|\Pi| \over z^2}. \eqno{(2.10)}$$
Therefore  the total energy diverges near the boundary of AdS$_3$, but it is on the other hand convergent near the event horizon. 

   We make furthermore  some comments which are of interesting.
 
    1.  The energy density in (2.9) is the energy per unit length $z$.  We can also calculate that of per unit length ($\ell$) of the AdS D1-brane. The result is 
$$ {\cal H_\ell}  = {|\Pi| \over \Pi^2 + z^4}. \eqno{(2.11)}$$
Therefore, we see that the energy per unit length of the AdS D1-brane approaches zero near the event horizon and approaches to the constant $1$ near the boundary of AdS$_3$.

   2.  Note that, the strings bounded  on D-brane are usually interpreted  as turning on the electric field on D-brane, as the strings are fusing inside the D-brane worldsheet by converting itself into homogenous electric flux.  Thus the strings living on the D-brane is characterized by the non-vanishing DBI electric field.  Using this property, one attempts to conclude that the vanishing DBI electric field will imply no F-string.    This is, however, not the case in here as there is a NS-NS $B^{NS}$ field.   In fact, from (2.4) we can easily see that only if $ E = - R^2$ could the value $\Pi = 0$ and there is no string on the D1-brane.  This tells us that the DBI electric field of D-brane in the macroscopic strings background is not zero even if there is no string on it .

    3.  From (2.8) we can be seen that the term $- \Pi \, R^2$ is coming from the $B^{NS}$ field in the $AdS_3$ spacetime. Without the $B^{NS}$ field the $\Theta^z_{\; z}$ is definitely positive and there will not be able to exist the AdS D1-branes.   The investigation of this paper have shown that the $B^{NS}$ field plays a  crucial role in the existence of the AdS D-brane in the curved spacetimes. 

    4.  If we consider the case of straight D1-brane, then equation (2.6) tells us that the energy density becomes  ${\cal H} =\left( \sqrt{1+\Pi^2} - \Pi\right) \, R^2$.   Thus, the straight D1-brane will be contracted to the event horizon $R=0$ in which its energy is vanishing.  This configuration has a lower energy then that of the AdS D1-brane (see (2.11)). 

\subsection {AdS D2-branes in $AdS_3 \times S^3$ Background}
The background metric and NS-NS potential in the $AdS_3 \times S^3$ background  is 
$$ds^2= - r^2 \, dt^2 + r^2\,dz^2 + r^{-2}\,dr^2 + d\theta^2 +sin^2\theta(d\phi^2+ sin^2\phi\,\,d\xi^2)\, ,  \eqno{(2.12a)}$$ 
$$B^{NS}= ~r^2 \, dt \,\wedge \,dz. ~~ \hspace{2.5cm}\eqno{(2.12b)}$$ 
We will consider the axially symmetric D2-branes which are described by the  worldvolume ($t $,$z $,$\theta$) The geometry of D2-branes is described by the function $R(z)$ as that shown in figure 2.  The D2-branes we considered will carry a time-independent electric field $E$ and a magnetic field $B$ such that the DBI 2-form field strength is [11]
$$ {F}= E \, dt\wedge dz + B \, dz\wedge d\theta. \eqno{(2.13)}$$
Through the standard procedure, as those described in section 2-1, the Lagrangian is 
$${\cal L} = - \sqrt{R^4 + R'^2 - (R^2 + E)^2 + B^2\, R^2},\eqno{(2.14)} $$
and the associated energy density $\cal H$ and the stress tension $\Theta^z_{\; z}$ are
$$\Theta^z_{\; z} = { \sqrt{1 + \Pi^2}\, \, (R^{4} + B^2) \over \sqrt{R^4+ B^2\,R^2 +{R^\prime}^2 }} - \Pi \, R^2 , \eqno{(2.15)}$$ 
$${\cal H} =  \sqrt{1 + \Pi^2}\,  \sqrt{R^4+ B^2 \,R^2 + {R^\prime}^2 } - \Pi \, R^2,     \eqno{(2.16)}$$ 
in which $\Pi$ is proportional to the number of string and $B$ to the number of D0, which are living on the D2-brane. Using above results we can find following solutions.
\subsubsection{AdS D2-branes Without $B$ field}
From (2.14) we see that if $B=0$ then the Lagrangian of D2 becomes that of D1 in (2.3).  In this case, the geometry of the AdS D2-branes is described by a series of AdS D1-branes which are carrying a fixed value of $\Pi$ and array around the $z$ axes, as shown in figure 2. 
\\
\vspace{1cm}
\hfil\scalebox{1}{\includegraphics{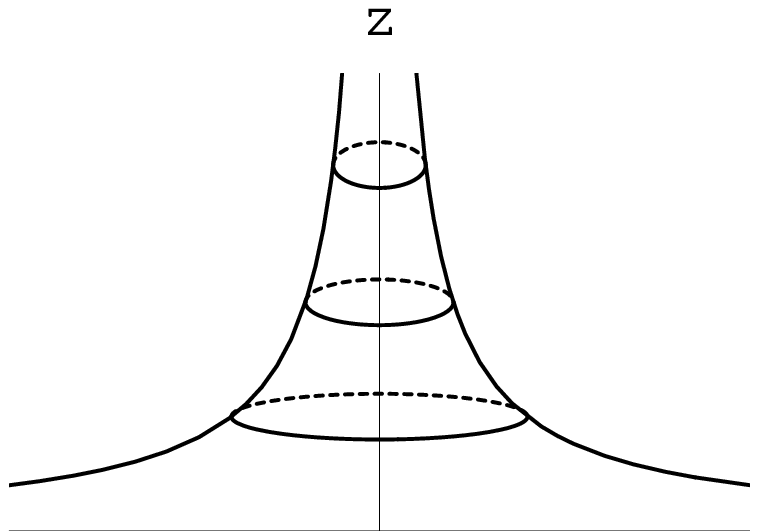}}\hfil\\
{\it ~~~Figure 2. The profile of the AdS D2-branes which are the axially symmetric bound states of a D2-brane with $n$ fundamental strings.}
\vspace{1cm}
\subsubsection{AdS D2-branes With $B$ field}
For the general case with finite DBI $B$ field we can follow the arguments in section 2-1 and let $\Theta^z_{\; z} = 0$, which then implies
$$R'(z)^2 + V(R) = 0,\eqno{(2.17)}$$ 
with 
$$ V(R) = -{~1~\over \Pi^2}\,\left(R^2+B^2\right)\,\left[R^2+\left(1+\Pi^2\right)B^2\right].\eqno{(2.18)}$$ 
Before numerically solving (2.17) we first make following useful analyses in the two limits.
$$R \rightarrow 0 : ~~~~{dR\over dz} \approx {\sqrt{1+\Pi^2}\, B^2\over {\Pi}}~~~\Rightarrow~~~ R(z) \approx {\sqrt{1+\Pi^2}\, B^2\over {\Pi}}\,(z - z_0) ,~~\eqno{(2.19a)}$$ 
$$~R \rightarrow \infty :~~~~~ {dR\over dz} \approx {R^2\over {\Pi}}~~~~~~~~~~~\Rightarrow~~~~ R(z) \approx {- {\Pi} \over (z - \tilde z_0)}\, ,~~~~~~~~~~~~~~~~~\eqno{(2.19b)}$$ 
in which $z_0$ and $\tilde z_0$ are integration constants.  Equation (2.19a) tells us that the AdS D2-branes will end on the event horizon at $z=z_0$ and (2.19b) tells us that the AdS D2-branes will extend to infinite (i.e. $R \rightarrow \infty$) at $z=\tilde z_0$, as shown in the figure 3 which is plotted with a help of numerical calculation.
\\
\vspace{1cm}
\hfil\scalebox{1}{\includegraphics{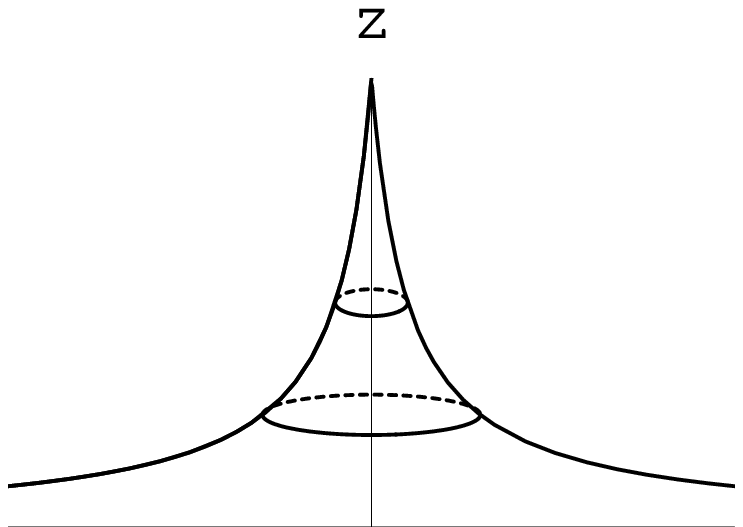}}\hfil\\
{\it ~~~Figure 3. The profile of the AdS D2-branes which are ending on the event horizon.}
\vspace{1cm}
\\
We make the following comments about the solution:

  1. The  pin-like geometry of the AdS D2-branes which end on the event horizon seems, more or less,  uncomfortable as the geometry of  D2-branes on the event horizon is singular.  This property can be easily seen from (2.19a) which shows that  the slope $R'(0)$ is finite.  However, as the coordinate system used to perform the analysis are not the global coordinates, and one cannot not infer from the analysis that the solution is singular.

  2.   Comparing to the fact that the AdS D2-branes which are not carrying the DBI $B$ field (i.e. D0 branes) will not end on the event horizon (as shown in section 2-2-1)  we may now say that the gravitational fields of the curved spacetime will attract the D0 branes (which are living on the ADS D-branes) and thus the AdS D2-branes to end on the event horizon.  This property could  also be found in the other curved spacetime, as shown in the section 4.

  3. In contrast to the AdS D1-branes (comment 1 in section 2-1), the energy density of the AdS D2-branes on the event horizon is nonvanishing, as can be seen by substituting the solution (2.19) into (2.16).   

  4.  From (2.16) we see that if $R'=0$ then ${\cal H}$ is an increasing function of $R$.   Thus, there does not exist any stable tubular solution with finite radius in the  $AdS_3 \times S^3$ background.  The configuration will collapse to the even horizon and has a vanishing energy.  The result is contrast to our previous investigations that the stable tubular bound states of ($nF1$, $m$D0, D2) could be formed in the NS5-brane and in the macroscopic strings background.  Note that the interesting problem that the entire $AdS_3 \times S^3$ geometry could be viewed as a supertube was  investigated in [18].

    5. The energy density of the solution of the AdS D2-brane configuration is 
$${\cal H} = {R'^2 \,\sqrt{1+\Pi^2}\over R^4+B^2\,R^2+R'^2},\eqno{(2.20)}$$
which is definitely positive if $R' \ne 0$.  Thus, comparing to those collapsing to the event horizon the AdS D2-branes have higher energy.
\section {AdS D-branes in NS5-branes Background}
The background fields around N  $NS5$-branes are given by the CHS solution [19]. The metric, dilaton and NS-NS $B^{NS}$ field are

$$ds^2=dx_\mu dx^\mu + h_{NS}(x^n) dx^mdx^m ,$$
$$e^{2(\Phi-\Phi_0)}=h_{NS}(x^n)\, ,~~~~~~~~~~~~~~~~~$$
$$H_{mnp} = -\epsilon_{mnp}^q\partial_q\Phi\,.~~~~~~~~~~~~~~~~ \eqno{(3.1)}$$
Here $h_{NS}(x^n)$ is the harmonic function describing $N$ fivebranes, and $H_{mnp}$ is the field strength of the NS-NS $B^{NS}$ field. For the case of coincident fivebranes one has
$$h_{NS}(r) =1 + {N \over r^2} , \eqno{(3.2)}$$
where $r = |\vec x|$ is the radial coordinate away from the fivebranes in the transverse space labeled by $(x^6,\cdots, x^9)$.  

   Using the above metric and fields we find that the Lagrangian and corresponding energy-momentum tensor of the static AdS D1-brane are
$$  L(R)  = -  \sqrt {\,h_{NS}^{-1} + \,R'^2 - h_{NS}^{-1} \, E^2}, ~~~~~~\eqno{(3.3)}$$
$$\Theta^z_{\; z} = {\sqrt{\,\Pi^2 \,+ h_{NS}^{-1}}\over \sqrt{\,1 + h_{NS}\, R'^2}}\, > 0 , \hspace{1.5cm}\eqno{(3.4)}$$
\\
respectively, in which $h_{NS}$ is defined by  (3.2) with $r \rightarrow R(z)$.  It can be seen that $\Theta^z_{\; z}$ is definitely positive and we does not have any stable  AdS D1-brane in the NS5-branes background. 

  In a same similar way, the Lagrangian and corresponding energy-momentum tensor the static AdS D2-brane are
$$  L(R)  = -  \sqrt {R^2 + \,h_{NS}^{-1}\, B^2 + h_{NS}\, R^2 \,R'^2 -  E^2\, R^2}, ~~~~~~\eqno{(3.5)}$$
$$\Theta^z_{\; z} = {\left(R^2 + h_{NS}^{-1}\,B^2 \right)\,\sqrt{\,\Pi^2 \,+ R^2}\over R\,\sqrt{R^2 + h_{NS}^{-1}\, B^2 + h_{NS}}}\, > 0 , \hspace{2cm}\eqno{(3.6)}$$
\\
respectively.  It can be seen that $\Theta^z_{\; z}$ is definitely positive and we still does not have any stable AdS D2-brane in the NS5-branes background. 

   Note that there are classical tubular D2-D0-F1 bound states in the NS5 background as found in our previous paper [12]. 

\section {AdS D-branes in Macroscopic Strings Background}
\subsection {AdS D1-branes in Macroscopic Strings Background}
The metric, the dilaton $(\phi)$ and the NS-NS $B^{NS}$ field  for a system of $N$ coincident macroscopic fundamental strings are given by 
$$ ds^2 = {1 \over h_f(r)} \Big( - dt^2 + dz^2 \Big) + dx^m dx^m , ~~~~~B^{NS}_{0z} =  h_f(r)^{-1} - 1 , ~~~~e^{2 \phi} =   h_f(r)^{-1},   \eqno{(4.1)}$$
where $ r$ denotes spatial coordinates transverse to the macroscopic string, $r \equiv {\sum x^m x^m}$, and the harmonic function $h_f({r})$ solving the  transverse Laplace equation is [20]
$$ h_f(r) = \Big( 1 + {N \over r^6} \Big). \eqno{(4.2)}$$
Using the above metric and fields we find that  the Lagrangian and corresponding energy-momentum tensor of the static AdS D1-branes are 
$$  L = -  \sqrt {\, h_f^{-1} + R'^2  -  h_f \left( E  +  h_f^{-1}-1\right)},~~~~~~~\eqno{(4.3)}$$
$$\Theta^z_{\; z} ={\sqrt{\,\Pi^2 \,+ h_f}\over h_f\,\sqrt{\,1 + h_f\,R'^2 }~} - \Pi\left(h_f^{-1}-1\right), \hspace{1.5cm}~\eqno{(4.4)}$$
$$~~{\cal H} = {\sqrt{(\,\Pi^2 \,h_f+ h_f^2)( h_f^{-1}+\,R'^2 )}\over h} - \Pi\left(h_f^{-1}-1\right). ~\eqno{(4.5)}$$
To find the function $R(z)$ we can follow the arguments in section 2-1 and let $\Theta^z_{\; z} = 0$, which then implies that
$$R'(z)^2 + V(R) = 0\, ,~~~~with~~~~~ V(R) = -~ {\Pi^2 + h_f \over \Pi^2\, h_f\,(1-h_f)^2} + h_f^{-1}\,.\eqno{(4.6)}$$ 
The equation (4.6) may be regarded as a particle with mass $2$ moving under the potential $V(R)$ while has zero total energy.   Therefore the variable $R$ represents the position of the particle and $z$ represents the ``time".  The potential $V(R)$  has two properties,
$$R \rightarrow 0 : ~~~~~~V(R)  \approx  {R^6 \over N},~~~~~~~~~~~\eqno{(4.7a)}$$ 
$$R \rightarrow \infty :~~~~~~V(R) \approx ~ - {1+\Pi^2\over N^2\, \Pi^2}\, R^{12},\eqno{(4.7b)}$$
as plotted in the figure 4. Thus, increasing $R$ from zero the potential will increase from zero and then drop to negative infinite, as shown in figure 4.   The particle coming from $R \rightarrow \infty$ will therefore hit on a turning point at finite value of $R_0$ at which $V(R_0) = 0 $. 

\vspace{1cm}
\hfil\scalebox{1}{\includegraphics{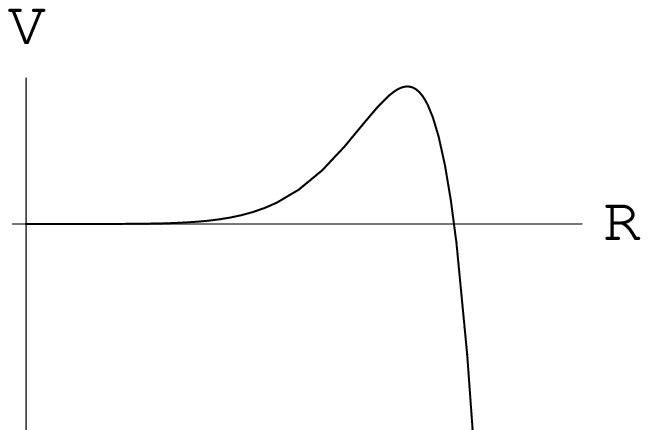}}\hfil\\
{\it ~~~Figure 4. The behaviors of the function $V(R)$.  There is a turning point at finite value of R.}
\vspace{1cm}
\\
Through calculation we find    
$$ R_0 = {N^{1/6}\,\Pi^{1/6}\over{\left(1 + \Pi^2\right)}^{1/6}}.\eqno{(4.8)}$$ Then it is useful to see the following analyses in the two limits.
$$R \rightarrow R_0 : ~~~{dR\over dz} \approx  - { 6\, \Pi ^{5/3}\, (R-R_0) \over N^{1/6}\,{\left(1 + \Pi^2\right)}^{1/6}}\,~~~\Rightarrow~~~ R(z) \approx R_0 \,+ \,exp\left[- { 6\, \Pi ^{5/3} \, (z -z_0)\over N^{1/6}\,{\left(1 + \Pi^2\right)}^{1/6}}\, \right]\,,~\eqno{(4.9a)}$$ 
$$~R \rightarrow \infty :~~~~{dR\over dz} \approx ~  {\sqrt{1+\Pi^2}\over N\, \Pi}\, R^{6}~~~~~~\Rightarrow~~~~ R(z) \approx \left[ {- 5\,\sqrt{1 + \Pi^2} \over N\, \Pi} \, (z - \tilde z_0)\right]^{-5},~~~~~~~~~~\eqno{(4.9b)}$$
in which $z_0$ and $\tilde z_0$ are the integration constants.  Equation (4.9a) tells us that the AdS D1-branes will asymptotically approach to the position ($R_0$) with a finite distance away from the macroscopic strings and (4.9b) tells us that the AdS D1-branes will extend to infinite (i.e. $R \rightarrow \infty$) at $z=\tilde z_0$.  The geometry of the AdS D1-brane is as that plotted in the figure 1 (in which ``d'' = $R_0$ ) with a help of numerical calculation.

   Now, if we consider the case of straight D1-brane, then the behavior of the energy density (4.5) shows that  the straight D1-brane will be contracted to the event horizon $R=0$.   From (4.5) we can see that ${\cal H}(R,R'\ne 0) \geq {\cal H}(R,R' = 0) \geq {\cal H}(R=R' = 0)$, thus the state of straight D1-brane collapsing to $R=0$ has a lower energy then that of the AdS D1-brane.  
\subsection {AdS D2-branes in Macroscopic Strings Background}
Through the standard procedure the Lagrangian and corresponding energy-momentum tensor of the static AdS D2-branes in macroscopic strings background are
$$  L = -  \sqrt {\, \left(h_f^{-1} + R'^2\right)\, R^2 + B^2 -  h_f \, R^2\,\left( E  +  h_f^{-1}-1\right)},~~~~~\eqno{(4.10)}$$
$$\Theta^z_{\; z} ={\sqrt{\,\Pi^2 \,+ R^2\,h_f}\,\left(R^2\,h_f^{-1} + B^2   \right)\over R\,\sqrt{\,R^2 + h_f\,R^2\,R'^2 + h_f\,B^2}~} - \Pi\left(h_f^{-1}-1\right). \hspace{1.5cm}~\eqno{(4.11)}$$
The energy density is
$$ {\cal H}(R) = {1\over R^6 + N}\sqrt{(R^8+(R^2\,R'^2+B^2 )(R^6+N B^2))(R^6+\Pi^2 R^4+N)} + {|\Pi| N\over R^6+N}.\eqno{(4.12)}$$
To find the function $R(z)$ we can as before let $\Theta^z_{\; z} = 0$, which then implies that $R'(z)^2 + V(R) = 0 $, in which 
$$V(R) = {\left(R^8 + B^2\,R^6+N\,B^2\right) \over R^2\,\left(R^6+N\right)} \left[\,1 - {\left(R^8 + B^2\,R^6+N\,B^2 \right)\, \left(R^6 + \Pi^2\,R^4+N\right) \over N^2\,\Pi^2}\right]\,.\eqno{(4.13)}$$ 
\\
The potential $V(R)$ has the following limits.
$$R \rightarrow 0 : ~~~ V(R) \approx {B^2\over R^2}\left[1- {B^2\over \Pi^2}\right]\,,~~~ or~~~ V(R) \approx {R^2\over N} ~~~ if~~ B=0,~\eqno{(4.14)}$$ 
$$R \rightarrow \infty :~~~~V(R) \approx - {R^{14}\over \Pi^2\,N^2}. \hspace{6.5cm}~~\eqno{(4.15)}$$
Then, depending on the values of $B$ and $\Pi$, the AdS d2-branes have two  types.

\subsubsection {AdS D2-branes with $|\Pi| > |B|$}
When $B=0$ the potential $V(R)$ will increase from zero and then drop to the negative infinity as shown in the dot line of figure 5 (or the figure 4).  When $|\Pi| > |B| > 0$ the potential $V(R)$ will decrease from the  positive infinity to the negative infinity as shown in the dash line of figure 5.  In these two cases, there is a turning point at a finite value of $R_0$.  Thus the axially symmetric bounded states have the geometry as those described in figure 2 while approach to a position with a finite distance away from the macroscopic strings.
\\
\\
\vspace{1cm}
\hfil\scalebox{1}{\includegraphics{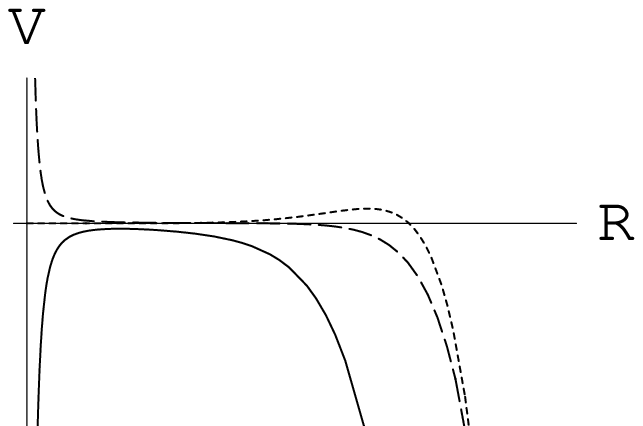}}\hfil\\
{\it ~~~Figure 5. The behaviors of the function $V(R)$.  The dot line describes  the potential in the case of $B=0$.  The dash line describes the potential in the case of $|\Pi |> |B| > 0$.  The solid line describes the potential in the case of $|B| > |\Pi|$.   }
\vspace{1cm}

\subsubsection {AdS D2-branes with $|B| > |\Pi|$}
On the other hand, when $|B| > |\Pi|$ then from (4.13) we can see that the potential $V(R)$ is definitely negative.  Thus, during increasing $R$ from zero  the corresponding potential will increase from the negative infinity to a maximum value which is still a negative value.   The potential will finally drop to the negative infinite, as could be seen from (4.15).  The corresponding $V(R)$ is plotted in the solid line of figure 5.

   From (4.14) and (4.15) we have the relations
$$R \rightarrow 0 : ~~~~{dR\over dz} \approx {B\over R}\sqrt{{B^2\over \Pi^2}-1 }\,~~~\Rightarrow~~~ R(z) \approx \sqrt{2B \left({B^2\over \Pi^2}-1\right)^{1/2}\,(z-z_0)} ,~~\eqno{(4.16)}$$ 
$$~R \rightarrow \infty :~~~~~ {dR\over dz} \approx {R^7\over N\,\Pi}~~~~~~~~~~~\Rightarrow~~~~ R(z) \approx \left({- N\Pi \over 6\,(z - \tilde z_0)}\right)^{1/6}\, ,~~~~~~~~~~~~~~~~~\eqno{(4.17)}$$ 
in which $z_0$ and $\tilde z_0$ are the integration constants.  Equation (4.16) tells us that the AdS D2-branes will end on the macroscopic strings at $z=z_0$ and (4.17) tells us that the AdS D2-branes will extend to infinite (i.e. $R \rightarrow \infty$) at $z=\tilde z_0$.  The geometry of the AdS D2-branes is therefore like that plotted in figure 3. 

  Comparing to the fact that the AdS D2-branes which are not carrying (or carrying a small) DBI $B$ field (i.e. D0 branes) will not end on the macroscopic strings, as that shown in section 4-2-1, we may say that the gravitational fields of the curved spacetime will attract the D0 branes (which are living on the ADS D-branes) and thus the AdS D2-branes to end on the event horizon.  This property has been found in the other curved spacetime, as shown in the section 2-2-2.

   Finally, it is worth to mention that, if we consider the case of cylindrically symmetric state then the corresponding energy density is 
$$ {\cal H}_{tube}(R) = {1\over R^6 + N}\sqrt{(R^8+B^2 R^6+N B^2)(R^6+\Pi^2 R^4+N)} + {|\Pi| N\over R^6+N},\eqno{(4.18)}$$
Using the relation ${\cal H'}(R_c) = 0$ we have the following nontrivial stable tube solutions with radius and energy density  
$$~~~~R_c = \sqrt{|B\Pi|} \,, ~~~~~ {\cal H}(R_c) = |B| + |\Pi| ,\eqno{(4.19)}$$
respectively, which was found by us in [12].  Now, comparing the (4.18) with (4.12) we see that
$${\cal H}(R) \geq {\cal H}_{tube}(R) \geq {\cal H}_{tube}(R_c).\eqno{(4.20)}$$ 
Thus the AdS D-2 branes found in this section have higher energy then  the tubular configurations.     

\section{Conclusion}
In this paper we use the Dirac-Born-Infeld action to find two classes AdS D-branes solutions in the $AdS_3 \times S^3$ and macroscopic strings backgrounds.  The first is the AdS D1-branes which are the bound states of a curved D1-brane with $n$ fundamental strings and the second is the axially symmetric bound states of a curved  D2-brane with $m$ D0-branes and $n$ fundamental strings 

   Through the detailed analyses, we have seen that, in the $AdS_3\times S^3$ background, the AdS D1-branes will asymptotically approach to the event horizon and the AdS D2-branes will end on the event horizon.  We have also show that there does not exist any stable AdS D-branes in the NS5-branes background.  For the case in the macroscopic fundamental string background,  we see that, both of  the AdS D1-branes and AdS D2-branes will keep a finite distance away from the macroscopic fundamental strings.   However, the AdS D2-branes could end on the macroscopic strings once it carries sufficient large D0-brane charges.    

    We have seen that the $B^{NS}$ field plays a  crucial role in the existence of the AdS D-brane in the curved spacetimes, as the NS-NS field could produce an exclusive force to bend the straight strings, which are living on the D1 or D2 branes brane, to produce the corresponding AdS D-branes.  It is also seen that the gravitational field in the curved spacetime has an effect to attract the D0 branes (which are living on the ADS D-branes) and thus the AdS D2-branes to end on the event horizon (or macroscopic fundamental strings), once it carries sufficient D0-branes charges. 

   We have also seen that AdS D1-branes and AdS D2-branes have higher energy then the configuration of the straight D1-brane and tube solution, respectively.   Finally, the problems of finding the supergravity solution and boundary states of the AdS D-branes in the NS5-branes and macroscopic F-strings backgrounds are of interesting and remain to be studied in the future. 
\\
\\
\\
\newpage
{\Large \bf  References}
\begin{enumerate}
\item D.~Berenstein, J.~M.~Maldacena and H.~Nastase, ``Strings in flat
space and pp waves from N = 4 super Yang Mills'',  JHEP {\bf 0204},
013 (2002) [hep-th/0202021];\\
 O.~Aharony, S.~S.~Gubser, J.~Maldacena, H.~Ooguri and Y.~Oz,  ``Large N field theories, string theory and gravity'',  Phys.\ Rept.\  {\bf 323} (2000) 183 [hep-th/9905111].
\item A.~Giveon, D.~Kutasov and N.~Seiberg, ``Comments on string theory on AdS$_3$", Adv.\ Theor.\ Math.\ Phys.\ {\bf 2}, 733 (1998) [hep-th/9806194];\\
J.~de Boer, H.~Ooguri, H.~Robins and J.~Tannenhauser, ``String theory on AdS$_3$", JHEP {\bf 9812}, 026 (1998) [hep-th/9812046];\\
D.~Kutasov and N.~Seiberg, ``More comments on string theory on AdS$_3$", 
JHEP {\bf 9904}, 008 (1999) [hep-th/9903219];\\
 J.~Maldacena and H.~Ooguri, ``Strings in AdS(3) and SL(2,R) WZW model. I'', J.\ Math.\ Phys.\  {\bf 42} (2001) 2929 [hep-th/0001053]; \\
 J.~Maldacena, H.~Ooguri and J.~Son, ``Strings in AdS(3) and the SL(2,R) WZW model. II: Euclidean black hole'', J.\ Math.\ Phys.\  {\bf 42} (2001) 2961 [hep-th/0005183];\\
 J.~Maldacena and H.~Ooguri, ``Strings in AdS(3) and the SL(2,R) WZW model. III: Correlation  functions'', Phys.\ Rev.\ D {\bf 65} (2002) 106006 [hep-th/0111180].
\item S.~Stanciu, ``D-branes in an AdS(3) background'', JHEP {\bf 9909} (1999) 028 [hep-th/9901122];\\
A.~Y.~Alekseev and V.~Schomerus, ``D-branes in the WZW model'', Phys.\ Rev.\ D {\bf 60} (1999) 061901 [hep-th/9812193];\\
G.\ Felder, J.\ Fr\"ohlich, J.Fuchs, C.Schweigert,  ``The geometry of WZW branes'', J.\ Geom.\ Phys. {\bf 34} (2000) 162-190 [hep-th/9909030].
\item C.~Bachas and M.~Petropoulos, ``Anti-de-Sitter D-branes'', JHEP
{\bf 0102}, 025 (2001) [hep-th/0012234].
\item P. M. Petropoulos, S. Ribault,  ``Some comments on Anti-de Sitter D-branes'' , JHEP 0107 (2001) 036 [hep-th/0105252];\\
A.~Giveon, D.~Kutasov and A.~Schwimmer, ``Comments on D-branes in AdS(3)'', Nucl.\ Phys.\ B {\bf 615}, 133 (2001) [hep-th/0106005]; \\
Y. Hikida and Y. Sugawara, ``Boundary States of D-branes in $AdS_3$ Based on Discrete Series'', Prog. Theor.Phys. 107 (2002) 1245-1266 [ hep-th/0107189];\\
A.~Rajaraman and M.~Rozali, ``Boundary states for D-branes in AdS(3),'' Phys.\ Rev.\ D {\bf 66}, 026006 (2002) [hep-th/0108001];\\
A. Parnachev and D. A. Sahakyan, ``Some remarks on D-branes in $AdS_3$'', JHEP 0110 (2001) 022 [hep-th/0109150];\\
P.~Lee, H.~Ooguri and J.~w.~Park, ``Boundary states for AdS(2) branes in AdS(3),'' Nucl.\ Phys.\ B {\bf 632}, 283 (2002)  [hep-th/0112188];\\
B.~Ponsot, V.~Schomerus and J.~Teschner, ``Branes in the Euclidean AdS(3),'' JHEP {\bf 0202}, 016 (2002) [hep-th/0112198]. 
\item K. Skenderis and M. Taylor, ``Branes in AdS and pp-wave spacetimes'',
 JHEP 0206 (2002) 025 [hep-th/0204054]. 
\item B. Ponsot and  S. Silva, ``Are there really any $AdS_2$ branes in the euclidean (or not) $AdS_3$?'',  Phys. Lett. B551 (2003) 173-177 [hep-th/0209084].
\item S.~Ribault, ``Two AdS(2) branes in the Euclidean AdS(3)'', JHEP 0305 (2003) 003 [hep-th/0210248];\\
C. Deliduman, ``$AdS_2$ D-Branes in Lorentzian $AdS_3$'',  Phys.Rev. D68 (2003) 066006 [hep-th/0211288].
\item  J.~Kumar and A.~Rajaraman, ``New supergravity solutions for branes in AdS(3) x S**3'',' Phys.\ Rev.\ D {\bf 67}, 125005 (2003) [hep-th/0212145]; ``Supergravity Solutions for $AdS_3 \times S^3$ branes'', Phys.Rev. D69 (2004) 105023 [hep-th/0310056]; ``Revisiting D-branes in $AdS_3 \times S^3$'', Phys.Rev. D70 (2004) 105002 [hep-th/0405024 ].
\item  D. Israel, ``D-branes in Lorentzian AdS(3)'', [hep-th/0502159].
\item D. Mateos and P. K. Townsend, ``Supertubes'', Phys. Rev. Lett. 87 (2001) 011602 [hep-th/0103030];\\
 R. Emparan, D. Mateos and P. K. Townsend, ``Supergravity Supertubes'', JHEP 0107 (2001) 011 [hep-th/0106012];\\
 D.~Mateos, S.~Ng and P.~K.~Townsend, ``Tachyons, supertubes and brane/anti-brane systems'', JHEP  0203 (2002) 016 [hep-th/0112054];\\
 M. Kruczenski, R. C. Myers, A. W. Peet, and D. J. Winters,``Aspects of supertubes'',  JHEP 0205 (2002) 017 [hep-th/0204103];\\
D. Bak, K. M. Lee, ``Noncommutative Supersymmetric Tubes'',  Phys. Lett. B509 (2001) 168 [hep-th/0103148];\\
D. Bak and S. W. Kim, ``Junction of Supersymmetric Tubes,'' Nucl. Phys.  B622 (2002) 95 [hep-th/0108207];\\
 D. Bak and A. Karch, ``Supersymmetric Brane-Antibrane Configurations,'' Nucl. Phys. B626 (2002) 165 [hep-th/011039];\\
 D. Bak and N. Ohta, ``Supersymmetric D2-anti-D2 String,'' Phys. Lett.  B527 (2002) 131 [hep-th/0112034].
\item Wung-Hong Huang, ``Tubular Solutions in NS5-brane, Dp-brane and Macroscopic String Background'' JHEP 0502 (2005) 061 [hep-th/0502023]; ``Tubular Solutions of Dirac-Born-Infeld Action on Dp-Brane Background'' Phys. Lett. B608 (2005) 244 [hep-th/0408213].
\item D.~Kutasov, ``D-brane dynamics  near NS5-brane,'' [hep-th/0405058]; ``A geometric interpretation of the open string tachyon,'' [hep-th/0408073]; \\
Y.~Nakayama, Y.~Sugawara and H.~Takayanagi, ``Boundary states for the rolling D-brane in NS5 background,'' JHEP  0407(2004) 020 [hep-th/0406173];\\
D.~A.~Sahakyan, ``Comments on D-brane dynamics near NS5-brane,''
JHEP 0410 (2004) 008 (2004) [hep-th/0408070];\\
J.~Kluson, ``Non-BPS D-brane near NS5-brane,'' JHEP  0411 (2004) 013 [hep-th/0409298]; ``Non-BPS Dp-brane in the background of NS5-brane on transverse R**3 x S**1,'' [hep-th/0411014];\\
S.~Thomas and J.~Ward, ``D-brane dynamics and NS5 rings,'' JHEP 0502 (2005) 015 [hep-th/0411130]; ``D-brane dynamics near compactified NS5-branes'' [hep-th/0501192].
\item  A.~Ghodsi and A.~E.~Mosaffa, ``D-brane Dynamics in RR Deformation of NS5-branes Background and Tachyon Cosmology'' [hep-th/0408015];\\
H. Yavartanoo, ``Cosmological Solution from D-brane motion in NS5-Branes background'' [hep-th/0407079];\\
B.~Chen, M.~Li and B.~Sun, ``D-brane near NS5-brane:  With electromagnetic field,''  JHEP 0412 (2004) 057 [hep-th/0412022];\\
Y.~Nakayama, K.~L.~Panigrahi, S.~J.~Rey and H.~Takayanagi, ``Rolling down the throat  in NS5-brane background: The case of electrified D-brane,'' JHEP 0501 (2005) 052 [hep-th/0412038].
\item K.~L.~Panigrahi, ``D-brane dynamics in Dp-brane background,'' Phys. Lett. B  601 (2004) 64 (2004) [hep-th/0407134];\\
O.~Saremi, L.~Kofman and A.~W.~Peet, ``Folding branes,'' [hep-th/0409092];\\
J.~Kluson, ``Non-BPS Dp-brane in Dk-Brane Background,'' [hep-th/0501010];\\
P. Bozhilov, `` Probe branes dynamics: exact solutions in general backgrounds", Nucl. Phys. B 656 (2003) 199 [hep-th/0211181];\\
C.~P.~Burgess, P.~Martineau,  F.~Quevedo and R.~Rabadan, ``Branonium,'' JHEP 0306 (2003) 037 [hep-th/0303170];\\
C.~P.~Burgess, N.~E.~Grandi,  F.~Quevedo and R.~Rabadan, ``D-brane chemistry,'' JHEP 0401 (2004) 067 [hep-th/0310010].
\item D.~Bak, S.~J.~Rey and H.~U.~Yee, ``Exactly soluble dynamics of (p,q) string near macroscopic fundamental strings,'' JHEP 0412 (2004) 008 [hep-th/0411099].
\item A.~Sen, ``Time and tachyon,'' Int.  J.  Mod.  Phys.  A 18 (2003) 4869 [hep-th/0209122];  ``Tachyon matter,'' JHEP  0207  (2002) 065 [hep-th/0203265]; ``Rolling tachyon,'' JHEP  0204 (2002) 048 [hep-th/0203211]; ``Tachyon dynamics in open string theory,'' [hep-th/0410103].
\item  O. Lunin and S. D. Mathur, ``Metric of the multiply wound rotating string,'' Nucl.Phys. B610 (2001) 49 [hep-th/0105136]; ``AdS/CFT duality and the black hole information paradox ,'' Nucl.Phys. B623 (2002) 342 [hep-th/0109154];\\
O. Lunin, S. D. Mathur, I.Y. Park and A. Saxena, ``Tachyon condensation and `bounce' in the D1-D5 system,''  Nucl.Phys. B679 (2004) 299 [hep-th/0304007].
\item C.~G.~.~Callan, J.~A.~Harvey and A.~Strominger, ``Supersymmetric string solitons,'' [hep-th/9112030].
\item A.~Dabholkar, G.~W.~Gibbons, J.~A.~Harvey and F.~Ruiz Ruiz, ``Superstrings And Solitons,'' Nucl.\ Phys.\ B  340 (1990) 33.
\end{enumerate}
\end{document}